# Atomistic simulation of phonon heat transport across metallic nanogaps

Yangyu Guo [1*], Christophe Adessi [1], Manuel Cobian [2], Samy Merabia [1†]

[1] *Institut Lumière Matière, Université Claude Bernard Lyon 1-CNRS, Université de Lyon, Villeurbanne 69622, France*

[2] *Laboratoire de Tribologie et Dynamique des Systèmes, École Centrale de Lyon-CNRS, Université de Lyon, Ecully 69134, France*

(Dated: May 25, 2022)

## Abstract

The understanding and modeling of the heat transport across nanometer and sub-nanometer gaps where the distinction between thermal radiation and conduction become blurred remains an open question. In this work, we present a three-dimensional atomistic simulation framework by combining the molecular dynamics (MD) and phonon non-equilibrium Green's function (NEGF) methods. The relaxed atomic configuration and interaction force constants of metallic nanogaps are generated from MD as inputs into harmonic phonon NEGF. Phonon tunneling across gold-gold and copper-copper nanogaps is quantified, and is shown to be a significant heat transport channel below gap size of 1nm. We demonstrate that lattice anharmonicity contributes to within 20-30 % of phonon tunneling depending on gap size, whereas electrostatic interactions turn out to have negligible effect for the small bias voltage typically used in experimental measurements. This work provides detailed information of the heat current spectrum and interprets the recent experimental determination of thermal conductance across gold-gold nanogaps. Our study contributes to deeper insight into heat transport in the extremely near-field regime, as well as hints for the future experimental investigation.

[*] yangyuhguo@gmail.com
[†] samy.merabia@univ-lyon1.fr

# 1. Introduction

When the distance between two solid objects drops down to few nanometers or even smaller, heat transport lies in the extremely near-field regime where the interplay of thermal radiation and conduction plays an important role [1,2]. Understanding the extremely near-field heat transport is of vital importance in many applications including the scanning tunneling microscopy [3,4], heat assisted magnetic recording [5,6], non-contact friction [7-9], thermal contact resistance [10,11] and so on. The two recent experimental reports of thermal conductance in this regime [12,13] are however controversial. Kloppstech *et al*. [12] observed giant thermal conductance of nanometer gaps while Cui *et al*. [13] found much smaller values below the detection limit of their probe. The underlying physical mechanism of heat transfer thus remains still an open question.

One possible mechanism to explain the giant experimental thermal conductance in the gold tip-surface nanogap [12] is the phonon tunneling across the nanogap. As two solid objects are sufficiently close to each other, the transfer of lattice vibration (phonons) may occur via the direct atomic interaction across the nanogap. There have been already some investigations of phonon tunneling across nanogaps by either harmonic or anharmonic theoretical approaches. The harmonic approaches were based on the Landauer's formalism of quantum transport, with the phonon transmission obtained through various methods: (*i*) phonon nonequilibrium Green's function (NEGF) formalism [14-18], (*ii*) elastic continuum model [19-22] or (*iii*) harmonic lattice dynamics [23,24]. The harmonic phonon NEGF simulations were mostly based on simplified one-dimensional (1D) formalism except in the very recent first-principles modeling of phonon transport across a Si-Si nanogap [18]. In addition, the nanogap configuration was usually not relaxed in both harmonic phonon NEGF and lattice dynamic methods except in the first-principles study [18]. The lattice anharmonicity was instead included in the recent molecular dynamics (MD) simulation of heat transport across Pt-Pt nanogaps [25]. However, the anharmonic effect on the thermal conductance of the nanogap has not been elucidated yet.

In this work, we present an atomistic modeling of phonon heat transport across metallic nanogaps by combining MD simulation and three-dimensional (3D) phonon NEGF formalism. The relaxed nanogap configuration and atomic interaction forces are generated from MD as inputs into the phonon NEGF. Furthermore, we extract and compare the spectral thermal conductance across the nanogap from both methods, which has not been reported hitherto to be authors' best knowledge. Thus we provide a robust computational framework to investigate the effects of anharmonicity and electrostatics. The remainder of the paper is organized as follows. The atomistic simulation methodology will be introduced in Section 2, and the results and discussions are given in Section 3, with the concluding remarks finally made in Section 4.

## 2. Methodology

We model the phonon heat transport across fcc (face-centered cubic) metal-metal nanogap, as shown in Figure 1(a). Our atomistic simulation framework combines both classical MD [26,27] and phonon NEGF method [28,29]. The MD includes all the order of anharmonicity automatically, whereas the harmonic phonon NEGF is adopted in this work due to too large computational cost of the anharmonic formalism [30]. The same empirical atomic interaction potential is employed in MD and NEGF, as will be explained later.

*2.1 Molecular dynamics simulation*

The classical NEMD (non-equilibrium MD) simulation is implemented in the open-source package LAMMPS [31]. We consider two types of metallic nanogaps: Au-Au and Cu-Cu gaps. The goal of study on Au-Au nanogap is to explain the recent experimental data of thermal conductance between the gold tip and gold substrate [12,13]. The Cu-Cu nanogap is also considered to investigate the effect of anharmonicity since copper is less anharmonic than Au based on the value of the Grüneisen parameter [32]. The 12-6 Lennard-Jones pairwise potential is adopted as seen in Figure 1(b), with the parameters for Au and Cu detailed in Ref. [33], which provide a good description of the surface and interface properties relevant to the situation in the present study. The cut-off radius in MD simulation is 12 Å for both metals, which is large enough to include the long-range atomic interaction [33]. We have also checked that the long-range interaction between metallic atoms is indeed captured in the Lennard-Jones potential by comparing to the Grimme's density functional dispersion correction [34], as detailed in Appendix A. The parallel plate-plate configuration is considered as shown in Figure 1(a), where periodic boundary conditions are used in the transverse direction. A very large cross-section of 16 uc × 16 uc is adopted for both Au and Cu gaps after careful independence verification. Here 1 uc denotes one cubic conventional unit cell of fcc metals with 4 atoms. In the transport direction, we have sandwiched the nanogap between hot and cold thermostats with the same length of 4 uc. Two fixed layers with the same length of 2 uc are imposed on both ends of the system. The number of atomic unit cells in the nanogap region are 6 and 7 respectively for Au gap and Cu gap, which are equally divided into both sides. These numbers are large enough to ensure that there is no direct interaction between one side of the nanogap and the thermostat on the other side. The gap size ($d$) is defined as the distance between the centers of the two surface atomic layers adjacent to the gap.

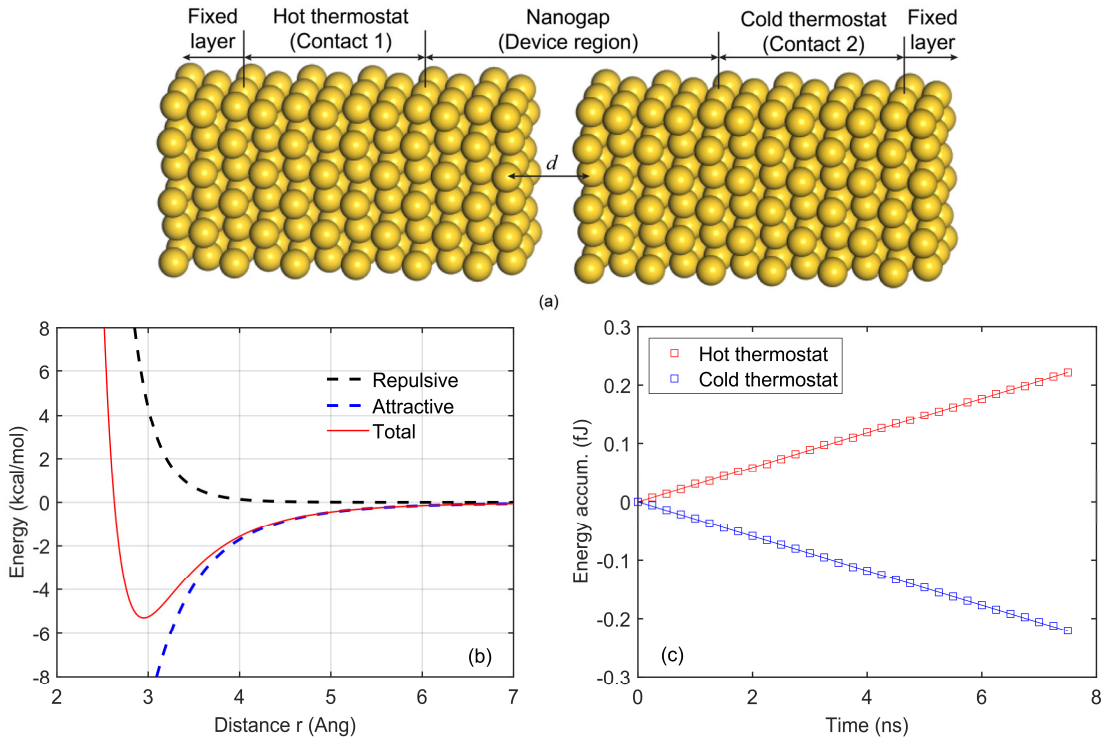

Figure 1. Atomistic simulation of phonon heat transport across metallic nanogap: (a) schematic of a Au-Au nanogap with a gap size $d$ and periodic cross-section; (b) Lennard-Jones potential for interaction between Au atoms; (c) energy accumulations at the hot and cold thermostats in MD (molecular dynamics) simulation of a Au-Au gap with $d$ = 3.94 Å at 300 K, the discrete points represent the MD data while the solid lines are linear fitting.

During the MD simulation, a time step of 0.5 fs is used. For both metal cases, firstly 1 million time steps (0.5 ns) are run to relax the whole system under *NPT* (isothermal-isobaric) ensemble. Then the fixed layers on both ends are fixed and 4 million time steps (2 ns) are run to make the free part reach a steady state under the effect of Langevin thermostats in the *NVE* (microcanonical) ensemble. Finally, 15 million time steps (7.5 ns) of steady-state runs are done for the calculation and analysis of the gap thermal conductance. To avoid the collapse of the gap structure due to the attractive interatomic force, a uniform tethering is exerted by attaching a harmonic spring (only along the transport direction) to each atom within the surface atomic mono-layers adjacent to the nanogap. The spring stiffness is low in order to not perturb the phonon dynamics too much, as to be discussed later. The thermal conductance of the nanogap is extracted based on the energy accumulation at the hot and cold thermostats, as exemplified in Figure 1(c) for a Au-Au gap with $d = 3.94$ Å at 300 K.

The frequency-dependent thermal conductance (or the spectral thermal conductance) is obtained by a spectral heat current (SHC) decomposition scheme [35,36], where the total heat current $Q$ is written as:

$$Q = \int_0^\infty q(\omega) \frac{d\omega}{2\pi} = \sum_{\substack{i \in I \\ j \in J}} \int_0^\infty q_{i \to j}(\omega) \frac{d\omega}{2\pi}, \qquad (1)$$

where $I$, $J$ denote separately the left and right sides of the nanogap region. The interatomic spectral heat current is calculated by: $q_{i \to j}(\omega) = 2\,\text{Re}\big[\tilde{K}_{ji}(\omega)\big]$, where 'Re' denotes the real part and $\tilde{K}_{ji}(\omega)$ is the Fourier transform of the atomic force-velocity correlation function defined as [35,36]:

$$K_{ji}(t_1 - t_2) = \frac{1}{2}\big\langle \mathbf{F}_{ji}(t_1) \cdot [\mathbf{v}_j(t_2) + \mathbf{v}_i(t_2)] \big\rangle. \qquad (2)$$

In Eq. (2), $\mathbf{F}_{ji}$ denotes the force acting on atom $j$ due to atom $i$. Following previous work, we consider only harmonic terms in $\mathbf{F}_{ji}$ such that the calculation of the spectral heat current is simplified [36]:

$$q_{i \to j}(\omega) = -\frac{2}{t_{simu}\,\omega} \sum_{\alpha,\beta} \text{Im}\big[\hat{v}_i^\alpha(\omega)^* \Phi_{ij}^{\alpha\beta} \hat{v}_j^\beta(\omega)\big], \qquad (3)$$

where $\hat{\mathbf{v}}_i(\omega)$ is the Fourier transform of the atomic velocity, $\Phi_{ij}^{\alpha\beta}$ is the harmonic (second-order) force constant matrix, 'Im' denotes the imaginary part, and $t_{simu}$ is the sampling simulation time. Note that the lattice anharmonicity is still included in the atomic velocity statistics in Eq. (3). In practical implementation, we output the time-dependent atomic velocities during the aforementioned 7.5 ns steady-state runs. The harmonic force constant matrix is computed based on the relaxed gap configuration obtained in the equilibration stage. The spectral thermal conductance is finally obtained as: $G_\omega = \dfrac{q(\omega)}{2\pi A_c \Delta T}$, with $A_c$ and $\Delta T$ the cross-sectional area and the temperature difference between thermostats respectively.

*2.2 Nonequilibrium Green's function method*

In parallel, the harmonic phonon NEGF formalism is adopted in this work, with the retarded Green's function calculated in matrix notation as [37,38]:

$$\mathbf{G}^R(\omega;\mathbf{q}_\perp) = \left[\omega^2 \mathbf{I} - \tilde{\mathbf{\Phi}}(\mathbf{q}_\perp) - \mathbf{\Sigma}^R(\omega;\mathbf{q}_\perp)\right]^{-1}, \tag{4}$$

where the superscript '-1' means the inverse of a matrix, $(\omega,\mathbf{q}_\perp)$ denote the frequency and wave-vector dependences along the transport direction and transverse periodic direction separately, $\mathbf{I}$ is the unity matrix, and $\tilde{\mathbf{\Phi}}(\mathbf{q}_\perp)$ is the Fourier's representation of the harmonic dynamic matrix [30,37]. The retarded self-energy matrix includes the contribution only from the two contacts: $\mathbf{\Sigma}^R(\omega;\mathbf{q}_\perp) = \mathbf{\Sigma}_1^R(\omega;\mathbf{q}_\perp) + \mathbf{\Sigma}_2^R(\omega;\mathbf{q}_\perp)$, which are related to the surface Green's function of the semi-infinite contacts as calculated by the decimation technique [39].

Once the retarded Green's function is obtained, the phonon transmission through the nanogap is calculated by [37,38]:

$$\Xi(\omega) = \frac{1}{N}\sum_{\mathbf{q}_\perp} \text{Tr}\left[\mathbf{\Gamma}_1(\omega;\mathbf{q}_\perp)\mathbf{G}^R(\omega;\mathbf{q}_\perp)\mathbf{\Gamma}_2(\omega;\mathbf{q}_\perp)\mathbf{G}^A(\omega;\mathbf{q}_\perp)\right], \tag{5}$$

where $N$ denotes the total number of discrete transverse wave vectors ($\mathbf{q}_\perp$), 'Tr' denotes the trace of a matrix, $\mathbf{G}^A$ the advanced Green's function as the Hermitian conjugate of $\mathbf{G}^R$, and the broadening matrices: $\mathbf{\Gamma}_{1(2)} = i\left[\mathbf{\Sigma}_{1(2)}^R - \mathbf{\Sigma}_{1(2)}^A\right]$, with $i = \sqrt{-1}$ here and $\mathbf{\Sigma}^A$ the advanced self-energy matrix as the Hermitian conjugate of $\mathbf{\Sigma}^R$. The thermal conductance is related to the transmission based on the Landauer's formula [28,29,37]:

$$G = \frac{1}{A_c} \int_0^\infty \hbar\omega \frac{\partial f_{BE}(\omega)}{\partial T} \Xi(\omega) \frac{d\omega}{2\pi}, \tag{6}$$

where $f_{BE}(\omega)$ is the Bose-Einstein distribution of phonons. To have a direct comparison to the classical MD simulation, the classical limit of Eq. (6) is written as follows by replacing the quantum heat capacity [$\hbar\omega \partial f_{BE}(\omega)/\partial T$] by the classical one ($k_B$):

$$G = \frac{1}{A_c} \int_0^\infty k_B \Xi(\omega) \frac{d\omega}{2\pi}. \tag{7}$$

As we aim to compare the phonon NEGF and MD results, the effect of tethering of the surface atomic monolayers adjacent to the nanogap in MD simulation must be also considered in phonon NEGF method. To do that, a diagonal term is added into the dynamic equation (4) of retarded Green's function:

$$\mathbf{G}^R(\omega; \mathbf{q}_\perp) = \left[ \omega^2 \mathbf{I} - \tilde{\mathbf{\Phi}}(\mathbf{q}_\perp) - \frac{k_{spring}}{m} \mathbf{I}_1 - \mathbf{\Sigma}^R(\omega; \mathbf{q}_\perp) \right]^{-1}, \tag{8}$$

where $\mathbf{I}_1$ is a diagonal matrix with diagonal components [1 0 0] associated with each atom under the tethering along only the transport direction (*x*-direction), and $k_{spring}$ the spring constant of the harmonic spring and *m* the atomic mass. Equation (8) is derived based on the lattice dynamic equation with an external tethering force, more details being given in Appendix B.

The required inputs into phonon NEGF include the atomic configuration of the nanogap and the harmonic force constant matrix, both of which are generated from MD simulation. A simulation cell with a cross-section of 1 uc × 1 uc is extracted from the cross-sectional center of the relaxed nanogap in the aforementioned equilibration stage. The length of the simulation cell for phonon NEGF is the same as that in MD simulation. We have checked that the transverse lattice symmetry is well preserved in the nanogap to justify the Fourier's representation with such a cross-section period in phonon NEGF. The harmonic force constant matrix of the nanogap is computed by combining LAMMPS with the same Lennard-Jones potential and the open-source package PHONOPY [40] based on the finite-displacement method. The phonon NEGF simulation is implemented in our highly-parallelized computational framework [30], with a very dense transverse wave vector of 100 × 100 after independence verification.

*2.3 Configurations of nanogaps*

In our NEMD simulation, the nanogaps will collapse in the natural state due to surface attraction. In the realistic situation of a tip-surface system, the tip is stretched as it approaches the surface due to attractive, adhesive forces between the tip and surface, as clearly shown in a previous experimental study [3]. We exert a tethering force on the surface atomic layers adjacent to nanogap to avoid the instability owing to structure relaxation. On the other hand, to minimize the influence of the tethering on the atomic and phonon dynamics around the nanogap, we use a tethering force as small as possible. In practical implementation, we construct a nanogap with an initial gap size $d_{\text{init}}$, and run *NPT* equilibrations under a series of decreasing tethering spring constants. We adopt the value of tethering spring constant just before the collapse of the nanogap. The relaxed gap size $d_{\text{relax}}$ under the adopted tethering is slightly smaller than the initial value ($d_{\text{init}}$). Several nanogaps are considered, with their initial (final) relaxed gap sizes and the corresponding tethering spring constants given in Table 1 and Table 2 for Au and Cu cases separately. Also we define the tethering spring frequency as $\omega_{\text{spring}} \equiv 2\pi f_{\text{spring}} = \sqrt{k_{\text{spring}}/m}$, which will be compared to the dominant frequency of phonons tunneled across the nanogap in later discussions.

Although the tethering is unavoidable, we will show that it does not perturb too much the phonon dynamics across the nanogap, especially at larger gap sizes. As the global variation of thermal conductance with nanogap size is much larger (several orders of magnitude), the tethering will not impact the general trend and final conclusion. Similar system collapse has been reported in a previous MD simulation of near-field heat transport between two silica nano-particles, where the authors stopped the simulation after a 10% reduction of the initial inter-particle distance [41]. Even though the stability issue was not explained in the recent MD study of heat transport across Pt-Pt nanogap [25], we believe the situation should be similar as the soft Lennard-Jones potential was also adopted.

Table 1. The initial and relaxed Au-Au nanogap configurations under the corresponding tethering spring of surface atomic layers adjacent to the gap in molecular dynamics simulation.

| $d_{init}$ (Å) | $k_{spring}$ (N/m) | $f_{spring}$ (THz) | $d_{relax}$ (Å) |
|---|---|---|---|
| 4.5 | 16.68 | 1.14 | 3.94 |
| 5.0 | 9.73 | 0.868 | 4.43 |
| 5.5 | 5.56 | 0.656 | 4.92 |
| 6.0 | 4.17 | 0.568 | 5.56 |
| 7.0 | 0.70 | 0.232 | 6.28 |

Table 2. The initial and relaxed Cu-Cu nanogap configurations under the corresponding tethering spring of surface atomic layers adjacent to the gap in molecular dynamics simulation.

| $d_{init}$ (Å) | $k_{spring}$ (N/m) | $f_{spring}$ (THz) | $d_{relax}$ (Å) |
|---|---|---|---|
| 4.0 | 18.07 | 2.08 | 3.45 |
| 4.5 | 11.12 | 1.63 | 4.05 |
| 5.0 | 6.95 | 1.29 | 4.62 |
| 5.5 | 4.17 | 1.00 | 5.12 |
| 6.0 | 2.78 | 0.82 | 5.67 |

*2.4 Electronic tunneling heat transport*

The electrons will tunnel across the metallic nanogap with a gap size below 1 nm [12,13] and also contribute to heat transport. For comparison, we estimate this electronic tunneling heat current based on the following formula:

$$Q_e = \int_0^\infty \left[ (E_x - \mu_1) N_1(E_x) - (E_x - \mu_2) N_2(E_x) \right] \tau(E_x) dE_x + \int_0^\infty dE_x \tau(E_x) \frac{m}{2\pi^2 \hbar^3} \int_0^\infty dE_r \left[ f_1(E) - f_2(E) \right] E_r,$$

(9)

where $E_x$ and $\tau(E_x)$ represent respectively the energy and the transmission probability of electrons along the transport direction, $E_r$ being the electron energy along the transverse direction in cylindrical coordinate, and the total electron energy $E = E_x + E_r$. The chemical potentials of electrons in the two metallic contacts are denoted by $\mu_1$ and $\mu_2$ respectively. The expressions of electron number distribution along the transport direction $N_{1(2)}(E_x)$ and Fermi-Dirac distribution $f_{1(2)}(E)$ in contacts are detailed in Appendix C, where the derivation of Eq. (9) is also shown. The present formulation is improved over the previous one in Ref. [17] in two aspects: (*i*) subtraction of the chemical potential in defining the heat current in contrast to an energy current in previous work; (*ii*) a more accurate account for the transverse kinetic energy ($E_r$) in contrast to an approximation ($k_B T$) in previous work.

## 3. Results and Discussions

In this section, we present the results of overall and spectral thermal conductance across the nanogaps by comparing the NEMD simulation and phonon NEGF modeling in Section 3.1. Then the effect of anharmonicity is investigated through the temperature-dependent results in Section 3.2. Finally the effect of electrostatics and a comparison with experimental data are discussed in Section 3.3 and Section 3.4 respectively.

### *3.1 Overall and spectral thermal conductance*

#### 3.1.1 Thermal conductance

The gap-size-dependent thermal conductances of the Au-Au and Cu-Cu nanogaps at room temperature (300 K) by NEMD and phonon NEGF are shown in Figure 2(a) and (b) respectively. In NEMD, we use a temperature difference of ± 50 K (i.e. 100 K), which is sufficiently small to remain in the linear regime while sufficiently large to have good statistics due to the very weak interaction across the nanogap. For both cases, we do not consider gap size larger than ~ 7 Å as the heat transport is too weak to obtain statistically meaningful heat currents in NEMD. On the other hand, we do not consider too small gap size as the nanogap is prone to collapse. In all, the nanogap size lies in a range where the atomic interaction is dominated by the attractive long-range dispersion force in Figure 1(b). The NEMD thermal conductance of nanogap decreases very rapidly with increasing gap size (*d*), following approximately a power law of $d^{-9}$. The usual NEGF conductance based on Eq. (4) is appreciably lower than the NEMD one at small gap size, which shall be mainly due to the perturbation stemming from the tethering treatment in NEMD. With the correction of the tethering force in Eq. (8), the NEGF conductance generally agrees with the NEMD one in the considered range of gap size. This also indicates the weak effect of anharmonicity at room temperature.

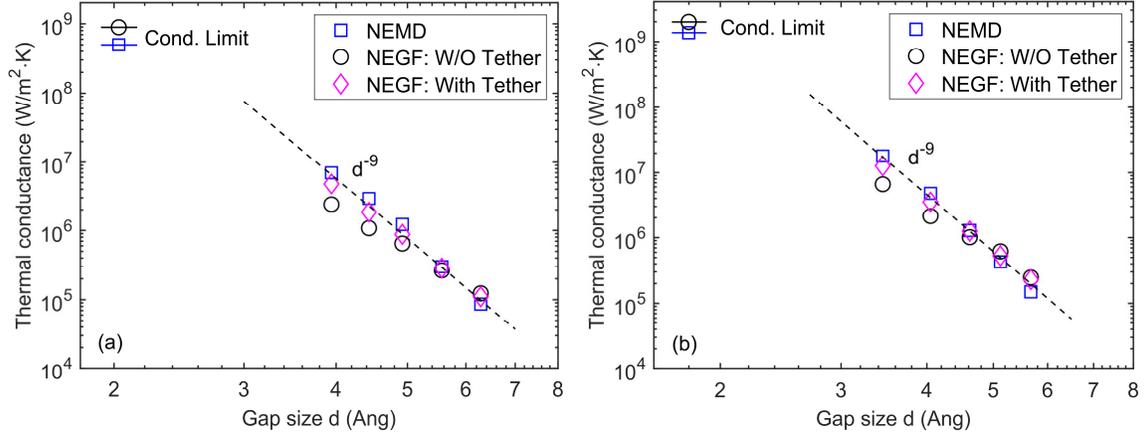

Figure 2. Thermal conductance versus gap size at room temperature (300 K): (a) Au-Au gap, (b) Cu-Cu gap. The blue squares denote the NEMD (non-equilibrium molecular dynamics) result, the magenta diamonds and the black circles represent the phonon NEGF (non-equilibrium Green's function) results with and without tethering of the surface atomic layers adjacent to the nanogap. The dashed line represents a power law scaling of $d^{-9}$. The conduction limit (Cond. Limit) corresponds to the contact situation (with a gap size of half lattice constant).

We now compare the calculated thermal conductance of Au-Au nanogap to the recent experimental results [12,13]. As the parallel plate-plate configuration is considered in our simulations, its thermal conductance per unit area ($G$) [W/m$^2 \cdot$K] is converted into the thermal conductance ($h$) [W/K] of a tip-plate configuration in experimental setup using the classical Derjaguin approximation [17,22]:

$$h(\tilde{d}) = \int_0^{r_{tip}} G(d) 2\pi r \mathrm{d}r , \qquad (10)$$

where $r_{tip}$ is the radius of the tip, and $r$ is the radius of a horizontal slice of the tip with a distance $d$ to the plate, with $d = \tilde{d} + r_{tip} - \sqrt{r_{tip}^2 - r^2}$, $\tilde{d}$ being the distance between the tip apex and the plate. Since the phonon thermal conductance $G(d)$ decreases very rapidly with $d$, the relevant gap size for the integration in Eq. (10) is much smaller than the tip radius in the experimental setup ($r_{tip}$ = 30 nm [12] and $r_{tip}$ = 150 nm [13]), such that $r_{tip} \gg d - \tilde{d}$ and we have approximately: $r^2 \simeq 2 r_{tip}(d - \tilde{d})$. Introducing $\tilde{d}' = (d - \tilde{d})$, we have $r^2 \simeq 2 r_{tip} \tilde{d}'$ and Eq. (10) is simplified into:

$$h(\tilde{d}) = 2\pi r_{tip} \int_0^{\frac{r_{tip}}{2}} G(\tilde{d}' + \tilde{d}) \mathrm{d}\tilde{d}' . \qquad (11)$$

In practical implementation, we consider the tip-plate gap size ($\tilde{d}$) up to 6 Å with the upper limit of the integration in Eq. (11) adopted as $\tilde{d}' = (9\text{ Å} - \tilde{d})$. This ensures the accuracy as the integrand $G$ at the upper limit ($d = 9$Å) is at least two orders of magnitude smaller than the value at the lower limit ($d = \tilde{d} \leq 6$ Å), as shown in Figure 3(a). As the NEGF result with tethering correction is very close to the NEMD one, we adopt the former as the thermal conductance of the plate-plate configuration. An analytical expression is obtained through a fitting of the phonon NEGF data in Figure 3(a), and then used for the integration in Eq. (11). For nanogap size ($d$) larger than those in Table 1, we follow the same procedure to generate the nanogap configuration and atomic interaction forces from MD as inputs into phonon NEGF. In Figure 3(a), the thermal conductance by the 1D phonon NEGF with Lennard-Jones potential [17] is also included, which shows almost the same power law scaling yet much larger magnitude compared to the present 3D NEGF. The underlying reason is that the average phonon group speed along different directions in 3D angular space is much smaller than that along mono-direction in the simplified 1D space [17]. This demonstrates the necessity of 3D atomistic simulation to accurately describe the phonon transport across vacuum nanogaps.

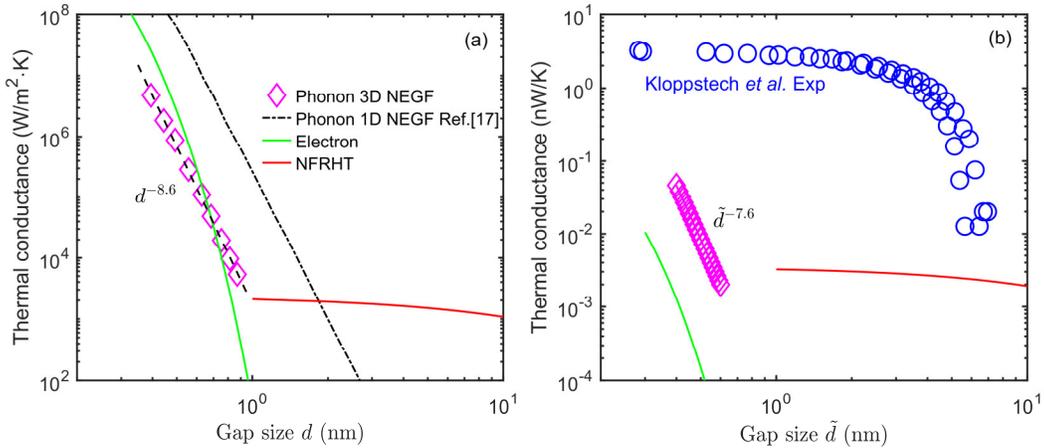

Figure 3. Thermal conductance versus gap size around room temperature: (a) plate-plate configuration, (b) tip-plate configuration corresponding to experimental setup. The magenta diamonds denote the present 3D phonon NEGF (non-equilibrium Green's function) result with tethering of the surface atomic layers adjacent to the nanogap, the black dash-dotted line represents the 1D phonon NEGF result from the reference [17], the red solid line denotes the NFRHT (near-field radiative heat transfer) result predicted by the fluctuational-electrodynamic theory, the green

solid line denotes the contribution of electron tunneling to heat transport, whereas the blue circles represent the experimental data [12].

The phonon thermal conductance of the gold tip-plate nanogap with a tip radius of $r_{tip}$ = 30 nm [12] is shown in Figure 3(b), with a power law scaling of $h \sim \tilde{d}^{-7.6}$. For comparison, we also include the electronic thermal conductance and the near-field radiative heat transfer (NFRHT) result by the fluctuational-electrodynamic theory [42] using the local dielectric function of gold [43]. The NFRHT thermal conductance is calculated for nanogap size only down to 1 nm below which the non-local effect will play a significant role and saturate the conductance [44]. The Derjaguin approximation in Eq. (10) is also valid for NFRHT, while not for electronic tunneling. We follow an approximate scheme for electron tunneling proposed in Ref. [22]: $h(\tilde{d}) = G(d = \tilde{d}) \cdot \pi r_{Au}^2$, with $r_{Au}$ = 1.35 Å the radius of a gold atom. As shown in Figure 3(b), the phonon thermal conductance is dominant over the photon and electron counterparts in this extremely near-field regime. However, the phonon thermal conductance is still much smaller than the experimental data of Kloppstech *et al.* [12]. With the tip radius of $r_{tip}$ = 150 nm [13], we obtain a maximal phonon thermal conductance of 0.23 nW/K at the minimal $\tilde{d}$ = 4 Å for the tip-plate configuration. This conductance is of the same order of magnitude as the measured maximal possible thermal conductance (~ 0.5 nW/K) of the smallest gaps after careful cleaning process [13]. The thermal conductance at larger gap size is indeed below the detection limit of their probe, as consistent with the conclusion drawn in the experiment of Cui *et al.* [13]. Further discussion about the explanation of the giant thermal conductance in the experiment of Kloppstech *et al.* [12] will be given in Section 3.4 later.

3.1.2 Spectral thermal conductance

The spectral thermal conductances of Au-Au nanogaps are extracted from NEMD simulations based on Eqs. (1)-(3), as shown in Figure 4(a). The phonon heat current spectrum is reduced very sharply beyond a critical frequency (~ 2 THz) as the higher-frequency phonons are more difficult to tunnel across the nanogap. With the increase of nanogap size from 3.94 Å to 6.28 Å, the spectral thermal conductance decreases about two orders of magnitudes, which explains well the rapid decay of thermal conductance in Figure 2(a). Such rapid decay of phonon tunneling with gap size is related to the $r^{-6}$ distance-dependent dispersion force of the Lennard-Jones potential. We also include the usual NEGF spectral conductance in Figure 4(a), which shows good consistency with NEMD result at high frequency while appreciable deviation at low frequency for the 3.94 Å and 4.92 Å nanogaps. The frequency range with large deviation is around the tethering spring frequency in Table 1, evidencing more definitely that the difference between the thermal conductances by NEGF and by NEMD at small gap size in Figure 2(a) is relevant to the perturbation introduced by tethering in NEMD. The perturbation becomes weaker with increasing gap size as the tethering force and spring frequency are smaller. With the tethering correction, the NEGF results of spectral thermal conductance show a general agreement with the NEMD ones, as illustrated in Figure 4(b). There is still some deviation around the tethering spring frequency, due to the fact that only the harmonic interaction is taken into account in NEGF.

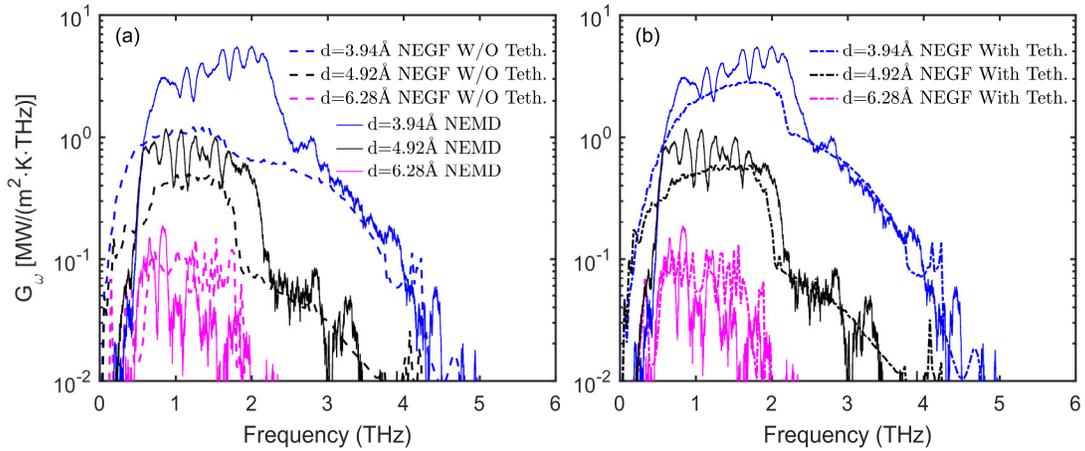

Figure 4. Spectral thermal conductance of Au-Au nanogaps at room temperature (300 K): NEMD result versus NEGF results (a) without and (b) with tethering of the surface atomic layers adjacent to the nanogap. The solid lines represent the NEMD (non-equilibrium molecular dynamics) results, whereas the dashed and dash-dotted lines represent the phonon NEGF (non-equilibrium Green's function) results without and with tethering respectively.

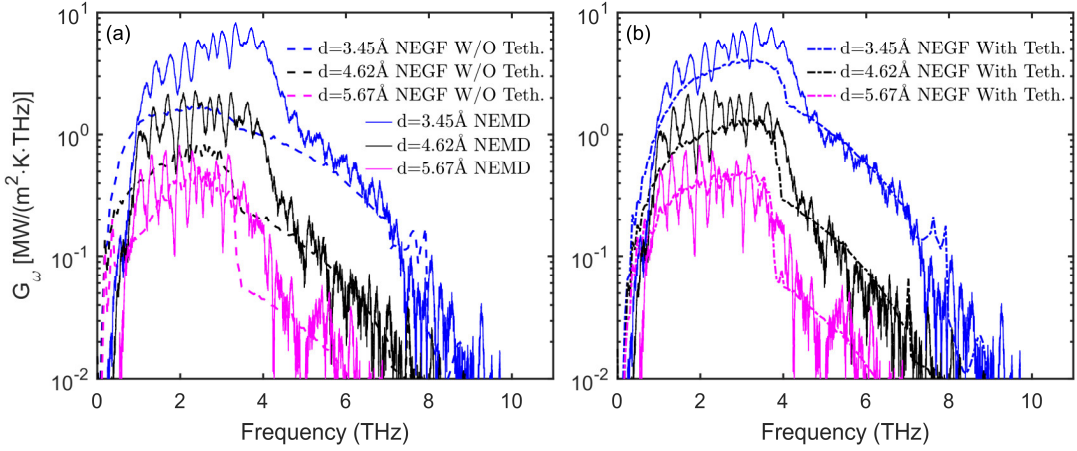

Figure 5. Spectral thermal conductance of Cu-Cu nanogaps at room temperature (300 K): NEMD result versus NEGF results (a) without and (b) with tethering of the surface atomic layers adjacent to the nanogap. The solid lines represent the NEMD (non-equilibrium molecular dynamics) results, whereas the dashed and dash-dotted lines represent the phonon NEGF (non-equilibrium Green's function) results without and with tethering respectively.

The results of Cu-Cu nanogaps are similar as shown in Figure 5. The critical frequency beyond which the phonon heat current spectrum is reduced sharply is almost twice (~ 4 THz) the value for the Au-Au nanogaps, as the atomic mass of Au is close to 3-4 times that of Cu. With the tethering correction, the deviation between NEGF and NEMD results around the tethering spring frequency is apparently smaller than that in Au-Au nanogaps. This could be explained by the weaker anharmonicity of Cu as compared to Au based on the average Grüneisen parameter of 2.0 and 3.0 respectively [32]. In addition, the spectral thermal conductance profile of Cu nanogap is more peaky than that of Au nanogap because of large fluctuations due to both smaller atomic mass and weak nature of atomic interaction across the nanogap.

*3.2 The effect of anharmonicity*

Anharmonic effects are known to play a non-negligible role in interfacial heat transfer [35,45] but remains ambiguous in heat transport across nanogaps. The present atomistic simulation framework provides a unique tool to uncover the role of anharmonicity through a comparison of the results of NEMD and harmonic phonon NEGF. We consider small Au-Au and Cu-Cu nanogaps with relaxed gap sizes of 3.94 Å and 3.45 Å respectively. In the Au nanogap, four system temperatures of 1 K, 10 K, 300 K and 500 K are considered, whereas in the Cu nanogap only the former three temperatures are considered. Note the classical statistics are considered in both NEMD and NEGF (via Eq. (7)), and the system temperature is varied to tune the strength of lattice anharmonicity. In NEMD simulations, the temperature differences of ± 0.5 K, ± 5 K, ± 50 K, ± 50 K are adopted at the four system temperatures respectively. Thermal expansion effect on the nanogap structure is shown to be very small by optimizing the lattice constant through *NPT* ensemble at each temperature. For instance, in the Au case, the lattice constants at the four temperatures are 4.0571 Å, 4.0571 Å, 4.0593 Å and 4.0611 Å respectively, whereas the relaxed nanogap sizes are 3.9455 Å, 3.9284 Å, 3.9429 Å and 3.9489 Å respectively.

The spectral thermal conductances of Au-Au nanogaps computed at different temperatures are shown in Figure 6. The NEGF spectral conductance with tethering correction is closer to the NEMD one at low temperatures (1 K and 10 K) in Figure 6 (a) and (b) compared to the situation at 300 K in Figure 6(c). This can be understood as the amplitude of atomic displacement is reduced with decreasing temperature such that the lattice vibration amplitude is small showing weaker anharmonicity. The ratio of thermal conductance by NEGF over that by NEMD is thus larger at 1 K than those at 10 K and 300 K, as shown in Figure 8(a). The NEMD thermal conductance is higher than the NEGF one (ratio < 1), as consistent with the trend of spectral thermal conductances. However, the complex interplay between tethering and anharmonicity makes the underlying mechanism still a bit elusive, which requires further study. As the temperature increases to 500 K, the enhanced anharmonic phonon scattering will suppress a bit heat transport in NEMD, such that the ratio of thermal conductances increases again. Note that the apparent agreement between NEMD spectral conductance and NEGF one with tethering correction in Figure 6 (d) is due to the consideration of only harmonic force terms in the SHC decomposition in Eq. (3). As shown in Figure 8(a), the thermal conductance obtained by NEGF is still around 20% lower than that obtained by the direct NEMD calculation at 500 K. The latter (namely, 6.41 MW/m$^2$·K) is slightly larger than the thermal conductance (6.06 MW/m$^2$·K) obtained by integrating the SHC from NEMD.

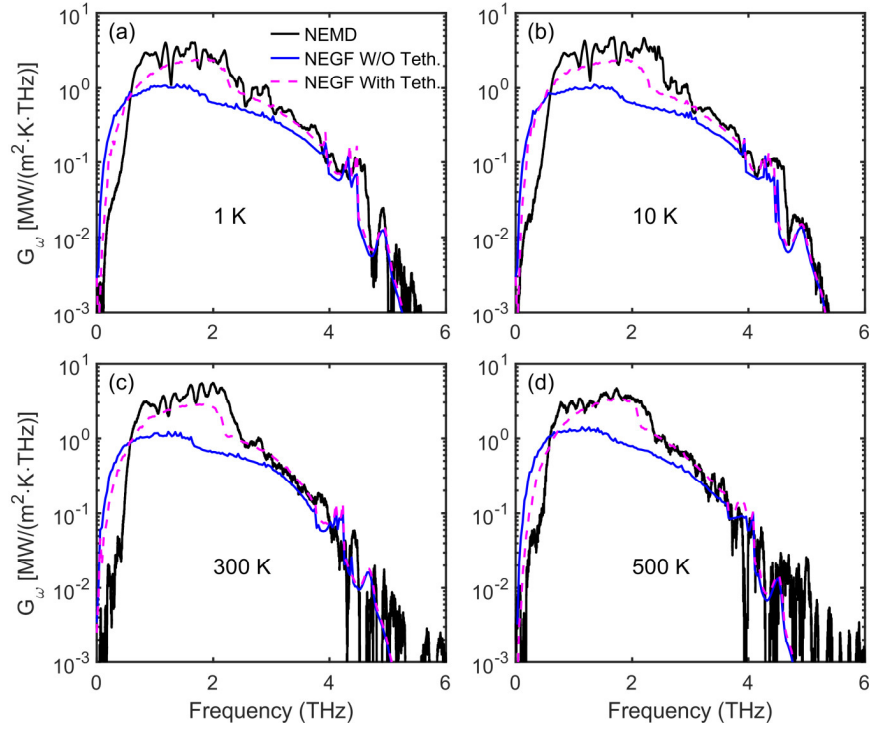

Figure 6. Spectral thermal conductance of Au-Au nanogap with $d$ = 3.94 Å at different temperatures: (a) 1 K, (b) 10 K, (c) 300 K, (d) 500 K. The black solid line represents the NEMD (non-equilibrium molecular dynamics) result, whereas the blue solid line and magenta dashed line represent the phonon NEGF (non-equilibrium Green's function) results without and with tethering of the surface atomic layers adjacent to the nanogap respectively.

The results for the Cu-Cu nanogaps are similar, as shown in Figure 7 and in Figure 8(b). As Cu is less anharmonic than Au, the spectral thermal conductance predicted by NEGF with tethering correction at lower temperatures (1 K and 10 K) is even closer to the NEMD result, as illustrated in Figure 7 (a) and (b). This is also reflected in the slightly larger ratio of thermal conductance by NEGF over that by NEMD as seen in Figure 8(b). In summary, the effect of anharmonicity is moderate in heat transport across Au-Au and Cu-Cu nanogaps. The difference between the anharmonic and harmonic approaches is around 20-30% in the studied temperature range. For larger nanogaps, the anharmonic effect shall be weaker as inferred from the small difference between the NEMD and NEGF thermal conductances in Figure 2.

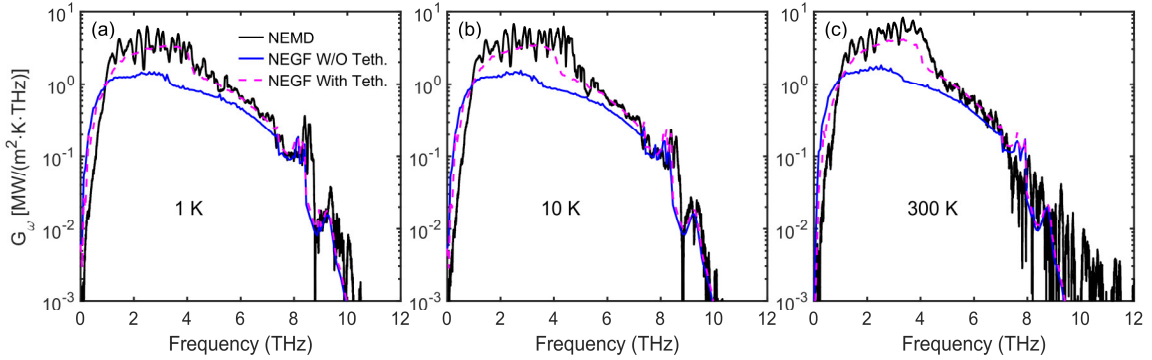

Figure 7. Spectral thermal conductance of Cu-Cu nanogap with $d$ = 3.45 Å at different temperatures: (a) 1 K, (b) 10 K, (c) 300 K. The black solid line represents the NEMD (non-equilibrium molecular dynamics) result, whereas the blue solid line and magenta dashed line represent the phonon NEGF (non-equilibrium Green's function) results without and with tethering of the surface atomic layers adjacent to the nanogap respectively.

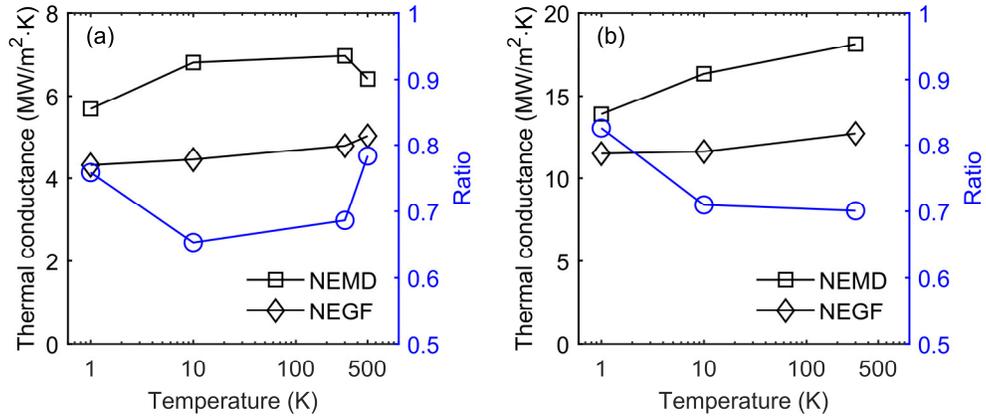

Figure 8. Temperature-dependent thermal conductance of nanogaps: (a) Au-Au gap with $d$ = 3.94 Å, (b) Cu-Cu gap with $d$ = 3.45 Å. The squares and diamonds with lines denote the NEMD result and NEGF result with tethering respectively, whereas the circles with line represent the ratio of the thermal conductance by NEGF over that by NEMD.

## 3.3 The effect of electrostatics

In the experimental measurements of the thermal conductance of nanogaps [12,13], a small bias voltage ($V$) is applied on the system for monitoring the tunneling current of electrons. Under the bias voltage, there will be induced surface charges which produce long-range Coulomb force. The Coulomb force may promote phonon tunneling through enhancing the atomic interactions across the nanogap. Here we explore this electrostatic effect by putting uniform charges with opposite signs on adjacent surfaces of the Au-Au nanogap at 300 K in the NEMD simulation. The surface charge is computed by $q_e = CV$, where the capacitance of the parallel plate-plate configuration is defined as $C = \varepsilon_0 A_c / d$, with $\varepsilon_0$ the vacuum permittivity. The capacitance and surface charge density is calculated based on the relaxed gap size ($d_{\text{relax, bare}}$) as in Table 1, whereas a voltage of 600 mV is adopted from the experimental report [12]. We re-optimize the nanogap configuration with surface charges under *NPT* ensemble and obtain a new relaxed gap size ($d_{\text{relax, charged}}$). In principle, the surface charge density has to be recalculated based on $d_{\text{relax, charged}}$ and the optimization is redone through a self-consistent iterative procedure. However, $d_{\text{relax, charged}}$ after the first iteration is quite close to $d_{\text{relax, bare}}$, as summarized in Table 3 for the different cases. Thus we adopt the nanogap configuration after the first iteration for further simulation and analysis.

Table 3. Initial and relaxed Au-Au nanogap configurations under the corresponding tethering without (bare) or with (charged) a bias voltage of 600 mV.

| $d_{\text{init}}$ (Å) | $d_{\text{relax,bare}}$ (Å) | Surface charge ($e$) | Surface charge per atom ($e$) | Surface charge density (C/m$^2$) | $d_{\text{relax, charged}}$ (Å) |
|---|---|---|---|---|---|
| 4.5 | 3.94 | 3.56 | $7.0 \times 10^{-3}$ | $1.35 \times 10^{-2}$ | 3.97 |
| 5.0 | 4.43 | 3.17 | $6.2 \times 10^{-3}$ | $1.20 \times 10^{-2}$ | 4.42 |
| 5.5 | 4.92 | 2.86 | $5.6 \times 10^{-3}$ | $1.08 \times 10^{-2}$ | 4.93 |
| 6.0 | 5.56 | 2.53 | $4.9 \times 10^{-3}$ | $9.6 \times 10^{-3}$ | 5.56 |
| 7.0 | 6.28 | 2.24 | $4.4 \times 10^{-3}$ | $8.5 \times 10^{-3}$ | 6.25 |
| 8.0 | 7.52 | 1.87 | $3.7 \times 10^{-3}$ | $7.1 \times 10^{-3}$ | 7.53 |

The thermal conductances of the bare and charged Au-Au nanogaps are shown in Figure 9, which show negligibly small difference. The results of large ($\geq 7.53$ Å) relaxed nanogaps are no longer shown as the thermal conductance is too low to obtain statistically meaningful heat current in NEMD even in the presence of surface charges. The spectral thermal conductances of three typical nanogaps are compared in Figure 10 between the bare and charged cases. Although there seems to be minor change of the heat current spectrum across the 6.28 Å nanogap induced by the surface charges, no significant effect is visible in the overall thermal conductance. The present conclusion is different from that of a recent 1D harmonic phonon NEGF modeling of Au-Au nanogap [17], which shows appreciable enhancement of the phonon thermal conductance by the bias voltage. The difference shall be mainly caused by the different treatments of the effect of electrostatic interaction. In Ref.

[17], an additional static surface charge was assumed on the sample surface following a previous experimental report on metal-dielectric tip-surface system [46]. Also the formula of total electrostatic force on the tip from both the dielectric surface and the metal electrode in Ref. [46] was adapted for the Coulombic interaction between the tip and surface [17]. Regarding the metal-metal tip-surface system considered here, we don't find clear evidence to assume an additional static surface charge on a metal surface as on a dielectric surface. Instead, we only consider the induced capacitive charges on the surfaces of Au-Au nanogap by the bias voltage.

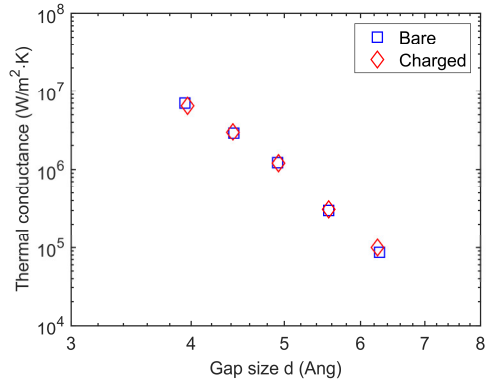

Figure 9. Electrostatic effect on thermal conductance of Au-Au gap via molecular dynamics at room temperature (300 K): the squares denote the result of bare nanogaps, whereas the diamonds denote the result of nanogaps with uniformly charged surfaces under a constant voltage of 600 mV, corresponding to the experimental setup of Kloppstech [12].

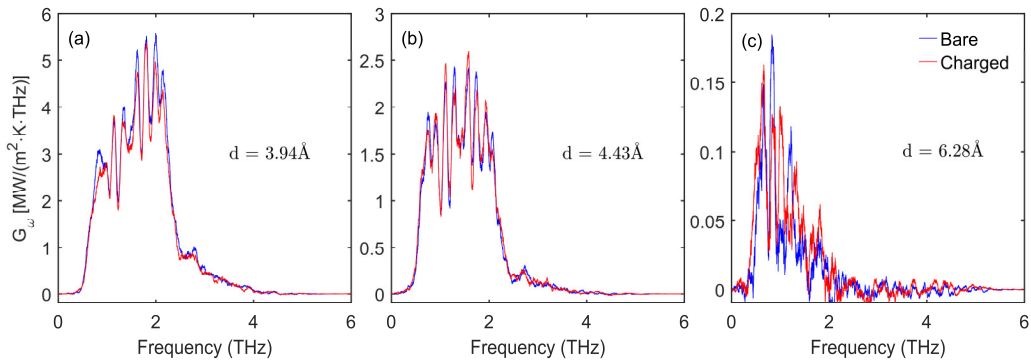

Figure 10. Electrostatic effect on spectral thermal conductance of Au-Au gap via molecular dynamics at room temperature (300 K): (a) $d$ = 3.94 Å, (b) $d$ = 4.43 Å, (c) $d$ = 6.28 Å. The blue solid lines denote the results of bare nanogaps, whereas the red solid lines denote the results of nanogaps with uniformly charged surfaces under a constant voltage of 600 mV, corresponding to the experimental setup of Kloppstech [12]. The relaxed gap sizes of the charged nanogaps are 3.97 Å, 4.42 Å and 6.25 Å respectively.

*3.4 Discussions*

One of the motivations of this work was to explain the experimental thermal conductance of nanogap in the extremely near-field regime by the phonon transport channel. Although our atomic simulation results are consistent with the experiment of Cui *et al.* [13], they are not able to reproduce the giant experimental conductance of Kloppstech *et al.* [12] in both the magnitude and the decaying slope as discussed in Section 3.1.1. Even including the large effect of the bias voltage in the electronic channel [22] as shown in Figure 11, we could not still explain the experimental data of Kloppstech *et al*. [12] especially beyond 1nm gap size. The heat current and heat flow are used here since the definition of electronic thermal conductance is not relevant in the presence of both temperature difference and chemical potential difference. We have also tried to model the phonon heat transport by NEGF with first-principles calculation of interatomic force constants across the nanogap, which are, however, comparable to or even smaller than the residual force toleration in structure relaxation for stable gaps larger than ~ 5 Å. Thus we could not obtain reliable first-principle modeling result. As all the possible channels (phonons, electrons, photons) have been considered, the disagreement is most probably due to the different conditions between the simulation and the experiment. Clean nanogaps with ideal surface conditions are modelled in the simulation. In contrast, there may be unknown contaminations from the sample fabrication and preparation procedure in experiment, as indicated by the cleaning-process-dependent apparent barrier heights and thermal conductances [13]. As no tunneling current was observed beyond 1nm gap size [12,13], the contamination shall be electrically insulating and is possibly organic molecules. Such organic molecules would bridge the tip and surface, and contribute to new heat transport channel across the nanogap. A gradual decrease of the thermal conductance has been observed due to the breaking of some molecular junctions when the tip is retracted from the surface within few nanometers [47]. This physical picture seems to be a possible explanation for the slow decay of thermal conductance versus gap size observed in the experiment of Kloppstech *et al.* [12]. To exclude the effect from contamination and achieve a more definite understanding of the underlying physics, further experimental investigation of the extremely near-field heat transport is pending in the future.

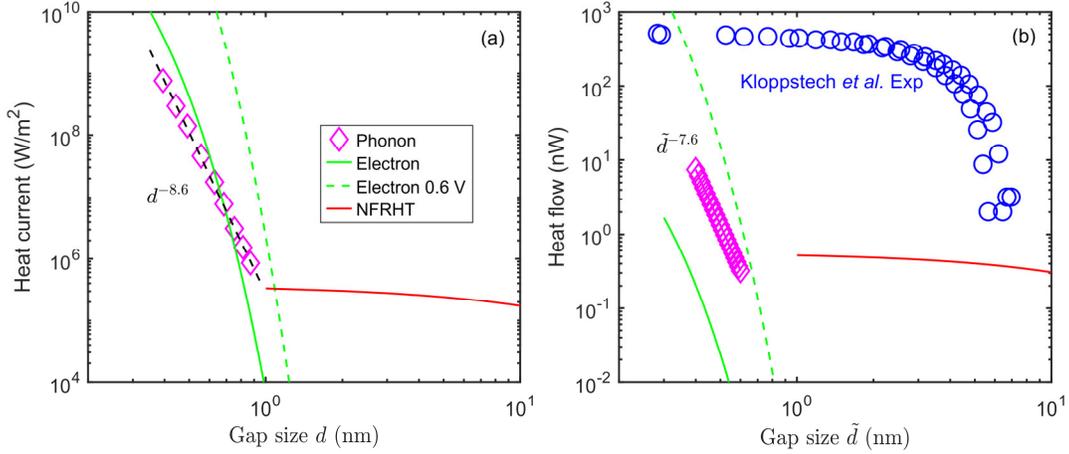

Figure 11. Heat transport across nanogaps around room temperature: (a) heat current in parallel plate-plate configuration, (b) heat flow in tip-plate configuration. The magenta diamonds denote the present 3D phonon NEGF (non-equilibrium Green's function) result with tethering of the surface atomic layers adjacent to the nanogap, the red solid line denotes the NFRHT (near-field radiative heat transfer) result predicted by the fluctuational-electrodynamic theory, the green solid line and dashed line denote respectively the contribution of electron tunneling to heat transport without or with a bias voltage, whereas the blue circles represent the experimental data [12].

## 4. Conclusions

In summary, we investigate the phonon heat transport across metallic nanogaps by combining molecular dynamics simulation and three-dimensional harmonic phonon non-equilibrium Green's function method. The tip-surface phonon thermal conductance decays rapidly with the gap size ($\tilde{d}$) through a power law of $\tilde{d}^{-7.6}$, and the phonon tunneling is a significant heat transport channel below the gap size of 1nm. The effect of lattice anharmonicity on phonon heat transport across nanogap amounts to within ~ 20-30% of the thermal conductance depending on gap size, whereas the impact of small bias voltage used in experimental measurement is found to be negligible. Our atomistic simulation is consistent with the experiment by Cui *et al*. [13] while not able to explain the giant heat transfer in the experiment by Kloppstech *et al.* [12] probably due to contaminations. The present atomistic simulation framework contributes to a more pertinent modeling of the phonon transport channel and hitherto detailed information of the heat current spectrum across nanogaps. This work thus promotes deeper understanding and future perspective of heat transport in the extremely near-field regime.

**Acknowledgements**

This work has been supported by the ANR project NearHeat (LCF Orsay and iLM Villeurbanne). This research used the computational resources of Raptor of iLM in Université Claude Bernard Lyon 1, the PSMN clusters in ENS de Lyon, and the Oakforest-PACs supercomputer system in The University of Tokyo. The authors appreciate helpful discussions with Dr. R. Messina & Dr. P. Ben-Abdallah from Université Paris-Saclay, Dr. A. Kyritsakis from University of Tartu, and Dr. S. Volz & Dr. Z. Zhang from The University of Tokyo.

**Appendix A. Comparison of Lennard-Jones potential and Grimme's dispersion correction**

To make sure that the Lennard-Jones potential captures the long-range dispersion force between metallic atoms, we compare to the Grimme's density functional dispersion correction (DFT-D3) [34]:

$$E_{disp} = -\frac{C_6}{r^6} f_{d,6}(r), \tag{A1}$$

where the $C_6$ coefficient is 342.3526 a.u. (atomic unit) for gold, and the damping function is defined as:

$$f_{d,6}(r) = \frac{1}{1 + 6\left[r/(s_{r,6} R_0)\right]^{-14}}. \tag{A2}$$

The atom pairwise cutoff radius for gold is: $R_0 = 3.26$Å, and the scaling factor $s_{r,6} = 1.532$ based on the PW6B95 functional is adopted due to its small mean absolute deviation [34]. The $C_8$ term in DFT-D3 is not considered here as it is more short-ranged. The dispersion correction in Eq. (A1) is shown in Figure 12 together with the Lennard-Jones potential used in the present work. The long-range attractive interaction term in Lennard-Jones potential has the same trend as the dispersion correction, and is negligibly small beyond 12 Å used as the cut-off radius in MD simulation.

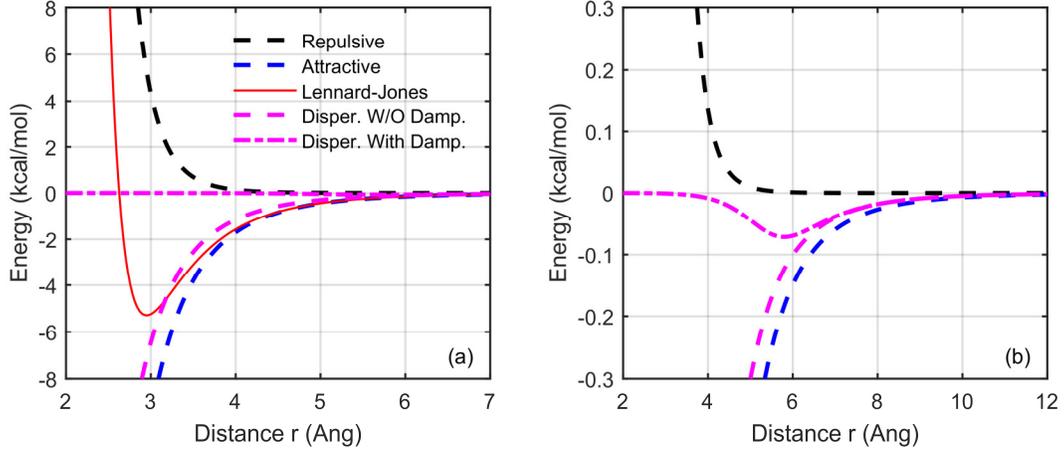

Figure 12. Comparison of Lennard-Jones potential and the Grimme's DFT-D3 dispersion correction for gold.

**Appendix B. Derivation of phonon NEGF formalism with external force**

Here we present a derivation of 1D phonon NEGF formalism with external tethering force, following the basic idea in classical works [28,48]. In the harmonic approximation, the lattice dynamic equation of atoms under an external harmonic spring is:

$$m_p \frac{d^2 u_p}{dt^2} = -\sum_q k_{pq} u_q - k_{spring} u_p,  \quad (B1)$$

where $u_p$ denotes the vibrational degree of freedom (i.e., atomic displacement) with $m_p$ the associated atomic mass, $k_{pq} = \partial^2 \Phi / \partial u_p \partial u_q$ is the second-order force constant, and $k_{spring}$ is the spring constant of the tethering spring.

Owing to the lack of translational lattice symmetry in the studied nanosystem by 1D phonon NEGF, the normal mode solution of Eq. (B1) has the following form [48]:

$$u_p = \frac{1}{\sqrt{m_p}} \phi_p \exp(-i\omega t),  \quad (B2)$$

with $\omega, \phi_p$ the frequency and amplitude of the vibrational degree of freedom respectively. Putting the solution of Eq. (B2) into the lattice dynamic equation in Eq. (B1), we obtain the dynamic equation in frequency domain as:

$$\left( \omega^2 - \frac{k_{spring}}{m_p} \right) \phi_p - \sum_q \Phi_{pq} \phi_q = 0,  \quad (B3)$$

where the components of dynamic matrix is defined as: $\Phi_{pq} \equiv k_{pq}/\sqrt{M_p M_q}$. Equation (B3) could be rewritten into the matrix notation as:

$$\left[\left(\omega^2 - \frac{k_{spring}}{m_p}\right)\mathbf{I} - \mathbf{\Phi}\right]\phi = 0, \tag{B4}$$

where $\mathbf{I}$ and $\mathbf{\Phi}$ are separately the unity matrix and dynamic matrix of dimension $N_p \times N_p$, whereas $\phi$ is a $N_p \times 1$ column vector, with $N_p$ the total number of vibrational degrees of freedom of the system. The retarded Green's function of the device region in the form of Eqs. (4) and (8) could be derived straightforwardly from the Eq. (B4) by splitting the system into two contact regions and a device region [48]. For the considered case with only tethering force along the transport direction (x-direction) in the present work, Eq. (B4) is reduced into:

$$\left[\omega^2 \mathbf{I} - \frac{k_{spring}}{m_p}\mathbf{I}_1 - \mathbf{\Phi}\right]\phi = 0, \tag{B5}$$

with the diagonal matrix $\mathbf{I}_1$ defined as:

$$\mathbf{I}_1 = \begin{bmatrix} 1 & & & & & & \mathbf{0} \\ & 0 & & & & & \\ & & 0 & & & & \\ & & & 1 & & & \\ & & & & 0 & & \\ & & & & & 0 & \\ \mathbf{0} & & & & & & \ddots \end{bmatrix}. \tag{B6}$$

Finally, the formulation of 3D phonon NEGF used in the present work [i.e., Eq. (8)] is obtained by introducing the Fourier's representation into the 1D formulation derived here along the transverse direction.

**Appendix C. Derivation of electronic tunneling heat current across vacuum nanogaps**

Here we derive the heat current associated with the electronic tunneling across a vacuum nanogap, following the idea of the derivation of electronic tunneling current [49]. Consider the two metallic contacts at temperatures $T_1$, $T_2$ and chemical potentials $\mu_1$, $\mu_2$ respectively. In the absence of a bias voltage, $\mu_1 = \mu_2 = E_f$, whereas in the presence of a bias voltage (V), $\mu_1 = E_f$ and $\mu_2 = E_f + eV$, with $E_f$ the Fermi level of the contact 1.

The heat current associated with the electronic tunneling from contact 1 to contact 2 is defined as:

$$Q_{e1} = 2\int_0^\infty \int_{-\infty}^\infty \int_{-\infty}^\infty v_x (E-\mu_1) f_1(E) \tau(E_x) \frac{dk_x dk_y dk_z}{(2\pi)^3}, \tag{C1}$$

where $v_x$, $E_x$ and $\tau(E_x)$ are the velocity, energy and transmission probability of electrons along the transport direction. The factor '2' expresses the spin degeneracy. The Fermi-Dirac distribution in contact 1 is expressed as $f_1(E) = \{\exp[(E-\mu_1)/k_B T_1] + 1\}^{-1}$. Assuming the free electron gas model for metals, we have the following parabolic dispersion relation [49]:

$$E = E_x + E_r = \frac{\hbar^2 k_x^2}{2m} + \frac{\hbar^2 k_r^2}{2m} = \frac{1}{2} m v_x^2 + \frac{1}{2} m v_r^2, \tag{C2}$$

where the cylindrical coordinate system is introduced for treating the periodic transverse direction: $k_r^2 = k_y^2 + k_z^2$, $v_r^2 = v_y^2 + v_z^2$. The integration in Eq. (C1) can be divided into two parts:

$$Q_{e1} = 2\int_0^\infty \int_{-\infty}^\infty \int_{-\infty}^\infty v_x (E_x - \mu_1) f_1(E) \tau(E_x) \frac{dk_x dk_y dk_z}{(2\pi)^3} + 2\int_0^\infty \int_{-\infty}^\infty \int_{-\infty}^\infty v_x E_r f_1(E) \tau(E_x) \frac{dk_x dk_y dk_z}{(2\pi)^3}. \tag{C3}$$

Integrating over the transverse wave vectors $(k_y, k_z)$ using the cylindrical coordinate system, we obtain from Eq. (C3):

$$Q_{e1} = \int_0^\infty (E_x - \mu_1) \tau(E_x) N_1(E_x) dE_x + \int_0^\infty dE_x \tau(E_x) \frac{m}{2\pi^2 \hbar^3} \int_0^\infty dE_r f_1(E) E_r, \tag{C4}$$

where $N_1(E_x)$ denotes the number of electrons in contact 1 per unit area per unit time per unit energy interval along the transport direction [50]:

$$N_1(E_x) = \frac{m k_B T_1}{2\pi^2 \hbar^3} \ln\left[\exp\left(-\frac{E_x - \mu_1}{k_B T_1}\right) + 1\right]. \tag{C5}$$

Through a similar procedure, the heat current associated with the electronic tunneling from contact 2 to contact 1 is derived as:

$$Q_{e2} = \int_0^\infty (E_x - \mu_2) \tau(E_x) N_2(E_x) dE_x + \int_0^\infty dE_x \tau(E_x) \frac{m}{2\pi^2 \hbar^3} \int_0^\infty dE_r f_2(E) E_r, \tag{C6}$$

where $f_2(E)$ and $N_2(E_x)$ are the counterparts in contact 2 of $f_1(E)$ and $N_1(E_x)$. Thus the net electronic tunneling heat current across the nanogap is calculated as $Q_e = Q_{e1} - Q_{e2}$:

$$Q_e = \int_0^\infty \left[(E_x - \mu_1)N_1(E_x) - (E_x - \mu_2)N_2(E_x)\right]\tau(E_x)dE_x + \int_0^\infty dE_x \tau(E_x)\frac{m}{2\pi^2\hbar^3}\int_0^\infty dE_r \left[f_1(E) - f_2(E)\right]E_r. \tag{C7}$$

The transmission probability of electrons across the vacuum nanogap is calculated based on the WKB approximation [49]:

$$\tau(E_x) = \exp\left[-\frac{\sqrt{8m_e}}{\hbar}\int_{x_1}^{x_2}\sqrt{W(x) - E_x}\,dx\right], \tag{C8}$$

where $m_e$ is the electron mass, $W(x)$ is the potential barrier profile in the nanogap, and $x_1$, $x_2$ are zeros of $W(x) - E_x$. We include the effect of image charge in the potential barrier profile: $W(x) = W_{id}(x) + W_{ic}(x)$, where the ideal linear profile and the image charge correction are expressed respectively as [17,50,51]:

$$W_{id}(x) = \mu_1 + W_f + \frac{x}{d}eV, \tag{C9}$$

$$W_{ic}(x) = \frac{e^2}{16\pi\varepsilon_0 d}\left[-2\psi(1) + \psi\left(\frac{x}{d}\right) + \psi\left(1 - \frac{x}{d}\right)\right]. \tag{C10}$$

In Eq. (C9), $W_f$ is the work function of the metallic contact 1. In Eq. (C10), $\varepsilon_0$ is the vacuum permittivity, and $\psi(z)$ is the digamma function with respective to a variable $z$.